\documentclass[apj]{emulateapj}
\usepackage{apjfonts}
\usepackage{graphicx}
\usepackage{color}
\usepackage{amsbsy}
\usepackage{mathrsfs}
\usepackage{mathpazo,bm}
\usepackage{amsmath}
\usepackage{multirow}

\newlength{\hfwidth}
\newlength{\hfwidthsingle}
\newcommand{\pderiv}[2]{\frac{\partial{#1}}{\partial{#2}}}

\newcommand{\ttimes}[1]{10^{#1}}

\renewcommand{\v}[1]{{\boldsymbol{#1}}} 

\newcommand{\ksi}{\xi}

\newcommand{\del}{\v{\nabla}}

\newcommand{\Div}{\del\cdot}

\newcommand{\mearth}{\,$M_{\oplus}$}
\newcommand{\mearthp}{\,$M_{\oplus}$ }

\newcommand{\cv}{c_{_{V}}}

\newcommand{\Figure}[1]{Figure~\ref{#1}}

\definecolor{brown}{rgb}{0.42,0.24,0.07}
\definecolor{darkgreen}{rgb}{0.0,0.6,0.00}
\definecolor{purple}{rgb}{0.7,0.0,0.7}
\definecolor{black}{rgb}{0.0,0.0,0.0}

\shorttitle{Outward migration in evolutionary models}
\shortauthors{Lyra et al.}

\slugcomment{Draft version}

\begin{document}

\title{Orbital migration of low-mass planets in evolutionary radiative
models: Avoiding catastrophic infall}

\author{
Wladimir Lyra\altaffilmark{1},
Sijme-Jan Paardekooper\altaffilmark{2}, \&
Mordecai-Mark Mac Low\altaffilmark{1}}

\altaffiltext{1}{Department of Astrophysics, American Museum of
  Natural History, 79th Street at Central Park West, New York, NY,
  10024, USA; wlyra@amnh.org, mordecai@amnh.org}
\altaffiltext{2}{Department of Applied Mathematics and Theoretical Physics, University of Cambridge, Wilberforce Road, Cambridge CB3 OWA, UK}




\begin{abstract}
{Outward migration of low-mass planets has recently been
shown to be a possibility in non-barotropic disks. We examine the consequences of this 
result in evolutionary models of protoplanetary disks.
Planet migration occurs toward equilibrium radii with zero 
torque.  These radii themselves migrate
inwards because of viscous accretion and 
photoevaporation. We show that as the surface density and temperature 
fall the planet orbital migration and disk depletion timescales eventually 
become comparable, with the precise timing depending on the mass of the planet. 
When this occurs, the planet decouples from the equilibrium radius. At
this time, however, the gas surface density is 
already too low to drive substantial further migration. 
A higher mass planet, of 10\mearth, can open a gap during the late
evolution of the disk, and stops migrating. Low-mass planets, with 1 or 0.1
\mearthp, released beyond 1~AU in our models avoid migrating into the
star. Our results provide support for the reduced migration rates
adopted in recent planet population synthesis models.} 
\end{abstract}

\section{Introduction}
\label{sect:introduction}

The origin of planetary systems remains a major challenge to
astrophysical theory. Aside from the quandary of planet formation, 
planet survival is also a problem. Planet formation occurs when the
gas disk is still present, and by exchanging angular momentum with the
gas, planets start to migrate. This occurs
as the planet excites perturbations in the disk that, in turn, exert
torques on the planet. The asymmetry of these perturbations on either
side of the planet determines the strength of the torques, and thus the
direction of migration.

The main components of the excited perturbations are the one-armed spirals
launched at the Lindblad resonances and the librating material on horseshoe orbits in the planet's corotation region. 
For low-mass planets, with weak wakes, both the shape of the spiral wake and the 
resulting Lindblad torque can be treated by linear analysis under some
assumptions, particularly of local isothermality. The 
analytical prediction (\citealp{Goldreich,Ward,Tanaka}), confirmed by numerical simulations 
(\citealp{Nelson,D'Angelo,Bate}), is that the inner Lindblad resonances lead 
to positive torques, whereas the outer ones lead to negative
torques. The outer Lindblad resonances lie closer to the planet, and
thus produce stronger torques, so the planet 
migrates inwards. This migration mode, referred to as Type I migration, 
occurs on timescales between $\ttimes{4}$ and $\ttimes{5}$ yr. This is a serious problem 
for planet formation since these timescales are much shorter than the lifetimes of
disks ($\ttimes{6}$-$\ttimes{7}$ yr). Halting  or slowing Type I migration is imperative if planets 
are to survive at all. Indeed, planet population synthesis models (\citealp[e.g.,][]{Alibert,Mordasini})
have to assume a reduction factor for Type I migration of
30--1000 in order to match the observed distribution of planetary semimajor axes. 

\citet{Paardekooper06} made a major step toward the solution
of this problem. They found that when the locally isothermal
approximation usually assumed in the literature was relaxed, the
planets migrated outwards. This behavior was explained by \citet{Baruteau} 
and \citet{Paardekooper08} as
resulting from an entropy-related torque exerted by material on
horseshoe orbits in the corotation region. 
This mechanism operates in regions of the disk 
that have a negative entropy gradient and 
inefficient radiative cooling, where sustenance of the torque
requires  some
viscous and thermal diffusion (\citealp{Paardekooper08}). 
\citet{Kley08} and \citet{Kley09} showed that outward migration indeed occurs in
disks with realistic heating and cooling.

This outward migration remains rapid. Planets migrate toward the
outer disk, reach an equilibrium radius of zero torque, and stay put
thereafter. As emphasized by \citet{Paardekooper10}, slow inward
migration then occurs as disk evolution shifts the equilibrium radius
inwards. The situation becomes similar to that of gap-opening planets
(Type II migration), with the planet migrating in lockstep with the
gas as the gas accretes.

This scenario poses a subtle problem that we examine in this
Letter. If the planet is moving together with the gas as the disk
depletes, at some point in the evolution the disk may reach a
thermodynamic state such that inward migration resumes. This brings
the problem back to square one, because, if the planet is to survive,
the remaining disk lifetime must be shorter than the Type I migration
timescale in that evolutionary state. We examine this possibility
using one-dimensional evolutionary models of protoplanetary disks
including heating and cooling. We describe the model in the following
section, present our results in \S~3, and give concluding remarks in
\S~4.

\section{The model}
\label{sect:model}

\subsection{Gas evolution}

We consider non-irradiated disks evolving by viscous 
diffusion and photoevaporation (\citealp{Lynden-Bell,Lin}),
\begin{equation}
\pderiv{\varSigma}{t}=\frac{3}{r}\pderiv{}{r}\left[r^{1/2}\pderiv{\varSigma\nu{r^{1/2}}}{r}\right]-\dot{\varSigma}_{\rm w}(r,t),\label{eq:density}
\end{equation}
where $\varSigma$ is the surface density, $r$ 
is the radius, and $\nu$ is the effective viscosity. We take the
photoevaporation rate to be (\citealp{Veras,Mordasini})
\begin{equation}
\dot\varSigma_{\rm w}=\left\{\begin{array}{ll}
0,&\mbox{~for~$r<R_g$},\\
\dot{M}_{\rm{w}}/[2\pi(r_{\rm ext}-R_g)r],&\mbox{~for~}r\geq{R_g},\end{array}\right.
\end{equation}which is valid for external irradiation.

For temperature evolution, we use a model without shock heating (\citealp{Nakamoto})
\begin{equation}
2\sigma{T^4}=\tau_{\rm eff}\left(\frac{9}{4}\varSigma\nu\varOmega^2\right)+2\sigma{T_b^4},\label{eq:temperature}
\end{equation}
where $T$ and $T_b$ are the midplane and background temperatures, respectively, 
$\varOmega$ is the Keplerian frequency, and $\sigma$ is the Stefan--Boltzmann constant. We take
the effective optical depth at 
the midplane (\citealp{Hubeny,Kley08}):
\begin{equation}
\tau_{\rm eff}=\frac{3\tau}{8}+\frac{\sqrt{3}}{4}+\frac{1}{4\tau}.\label{eq:taueff}
\end{equation}
Equation~(\ref{eq:temperature}) states that the emerging flux is the result of an equilibrium 
between viscous heating, background radiation, and radiative cooling. The optical depth 
is $\tau=\kappa\varSigma/2$, and 
the opacities $\kappa$ are taken from \citet{Bell}. We assume that although 
dust growth and planet formation lock away refractory material, 
fragmentation efficiently replenishes small grains, keeping 
the disks opaque during their evolution (\citealp{Birnstiel}).

\subsection{Planet evolution}

The planet's orbital radius evolves as 
\begin{equation}
\frac{d{r_p}}{dt}=\frac{2\Gamma}{m_pr_p\varOmega_p},\label{eq:planet}
\end{equation}where $m_p$ is the planet's mass and $\Gamma$ is the torque
from the gas. We assume a circular orbit and constant planet mass.
We only consider low-mass planets, and so can ignore the back reaction of the planet onto 
the gas (\citealp{Alexander}). The torques 
are modeled with analytical fits 
valid for the fully unsaturated case (\citealp{Paardekooper10}). Using $b/h$=0.4, where $h$ 
is the disk's aspect ratio and $b$ is the gravitational softening of the 
planet's potential (necessary in models with less than three dimensions), the torques are 
\begin{eqnarray}
\Gamma_{\rm{iso}}/\Gamma_0&=&-0.85-\alpha-0.9\beta,\\
\gamma\Gamma_{\rm{ad}}/\Gamma_0&=&-0.85-\alpha-1.7\beta+7.9\ksi/\gamma
\end{eqnarray}for the locally isothermal and adiabatic equations of state, 
respectively. The adiabatic index $\gamma=1.4$, and $\alpha$, 
$\beta$, and $\ksi$ are the negative of the local density, temperature, and 
entropy gradients:
\begin{equation}
\alpha=-\pderiv{\ln\varSigma}{\ln{r}};\qquad\beta=-\pderiv{\ln{T}}{\ln{r}};\qquad\ksi=\beta-(\gamma-1)\alpha.
\end{equation}
The torques are normalized by
\begin{equation}
\Gamma_0=(q/h)^2\varSigma_pr_p^4\varOmega_p^2,\label{eq:torque-norm}
\end{equation}
where $q$ is the ratio of planetary to stellar mass and $\varSigma_p$
is the surface density at the position of the planet.  We interpolate between the
two torque regimes to get
\begin{equation}
\Gamma=\frac{\Gamma_{\rm ad}\varTheta^2+\Gamma_{\rm{iso}}}{(\varTheta+1)^2}.\label{eq:torque}
\end{equation}

Here, $\varTheta=t_{\rm{rad}}/t_{\rm{dyn}}$, where $t_{\rm{rad}}$ and $t_{\rm{dyn}}$ are 
the radiative and dynamical timescales. If $\varTheta\ll{1}$, radiative processes 
can restore the temperature 
quickly compared to the horseshoe turnover time, 
leading to isothermal horseshoe turns. It was shown in \citet{Paardekooper09} that the 
timescale to establish the full horseshoe drag ($t_{\rm{drag}}$) is a fraction of the 
libration timescale (typically 10 $t_{\rm{dyn}}$ for a 5 \mearthp planet). However, 
as long as $t_{\rm{rad}} > t_{\rm{dyn}}$, the torque is affected by nonlinear effects, 
and since the nonlinear torque is so much stronger than its linear counterpart 
(\citealp{Paardekooper10}), the parameter governing linearity (and therefore 
isothermality) is $\varTheta$ rather than
$t_{\rm{rad}}/t_{\rm{drag}}$.

Since $t_{\rm{drag}}$ depends on $q$, using $t_{\rm{drag}}$ would lead to different 
migration behavior for different planet masses. In practice, the transition between 
isothermal and adiabatic regions occurs on such a small length scale that this would 
not change the qualitative outcome of the models. Mass segregation is more likely to 
come from effects of saturation, which we do not consider in this Letter, since it is 
as yet poorly understood for non-barotropic disks. We note, however, that for viscously 
heated, optically thick disks close to thermal equilibrium, 
radiative thermal diffusion will approach the viscous heating rate. Therefore, even in the 
very optically thick inner regions of the disk, a sizeable fraction of the unsaturated 
torque can be sustained as long as the disk remains viscous.

To calculate $\varTheta$, we consider $t_{\rm rad}=E/\dot{E}$, where $E$ is 
the internal energy. The cooling $\dot{E}  =\Div{\v{F}}$, where $\v{F}$ is 
the flux. Using $E=\cv\rho{T}$, $\rho=\varSigma/2H$, and 
$|\v{F}|=\sigma{T}_{\rm eff}^4$, we have

\begin{equation}
t_{\rm rad}=\frac{\cv\varSigma\tau_{\rm eff}}{6\sigma{T^3}}.
\end{equation}
The dynamical time $t_{\rm dyn}=2\pi/\varOmega$, so 
\begin{equation}
  \varTheta=\frac{\cv\varSigma\varOmega\tau_{\rm eff}}{12\pi\sigma{T}^3}.
\end{equation}

\subsection{Simulation parameters}

We use a one-dimensional linear grid covering 0.1--30~AU with 200
points.  The surface density is specified by the initial mass
accretion rate $\dot{M}_0$, and the viscosity parameter 
$\alpha_{\rm SS}$ (\citealp{Shakura}), following the analytical fits of \citet{Papaloizou}
for $\varSigma$--$\nu$ relations.  We use
$\dot{M}_0=\ttimes{-7}$\,M$_{\odot}$~yr$^{-1}$ and $\alpha_{\rm
  SS}=\ttimes{-2}$. The wind is modeled with $\dot{M}_{\rm
  w}=\ttimes{-8}$\,M$_{\odot}$~yr$^{-1}$ and $R_g=5$\,AU.  For a
given surface density, 
Equation~(\ref{eq:temperature}) specifies the temperature, with $T_b=10$\,K.  Because the optical depth
depends on temperature, we solve Equation~(\ref{eq:temperature}) with a
Newton--Raphson root-finding algorithm (using 0.01\,K precision). We
examine planets of mass 0.1, 1, and 10\mearth.  Planet--planet
interactions are ignored. Boundary conditions are taken as outflow. We
compute the derivatives as in the {\sc Pencil Code}{\footnote{See
    http://www.nordita.org/software/pencil-code}}, with sixth-order
spatial derivatives and a third-order Runge--Kutta time integrator.

\section{Results}

\begin{figure*}
\begin{center}
  \resizebox{.9\textwidth}{!}{\includegraphics{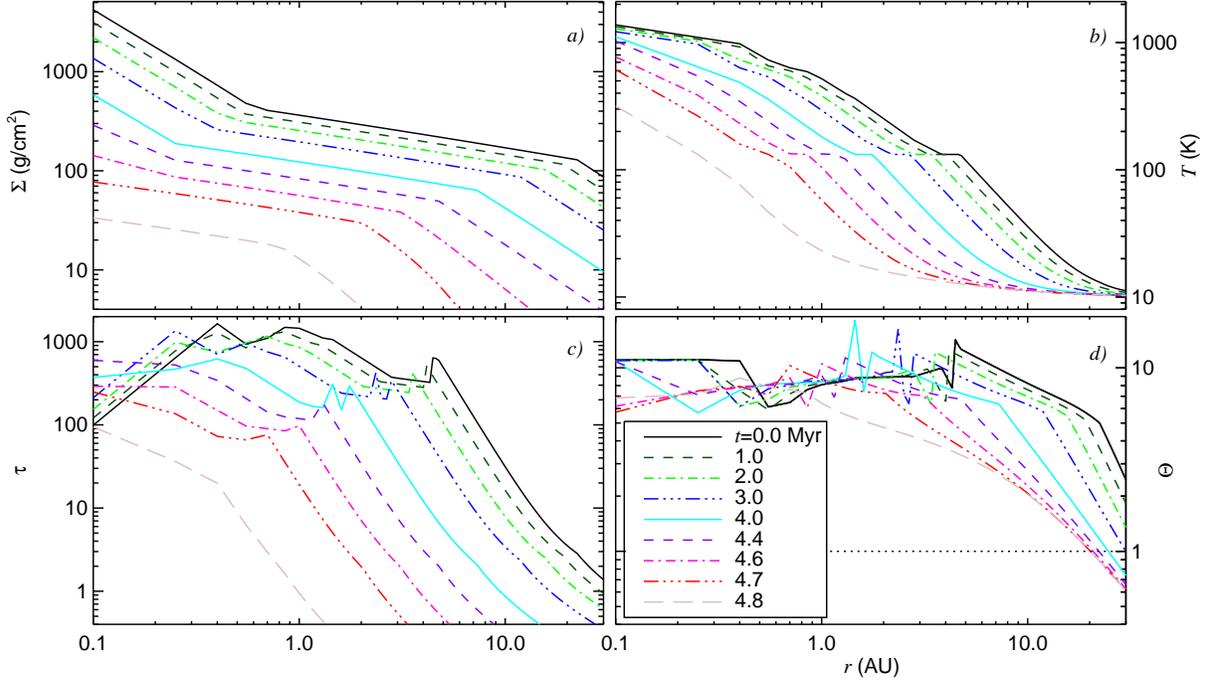}}
\end{center}
\caption[]{Evolution of (a) surface density,  (b) midplane temperature, 
 (c) optical depth, and (d) $\varTheta$, the ratio of radiative to dynamical 
timescales. The disk is drained after 4.8\,Myr, with mass dropping
below \mearth. 
The temperature shows a plateau at $\approx$130\,K, produced by the opacity 
transition when ice grains sublimate that moves inward as the disk cools. 
Through most of the evolution, $\varTheta>1$, so torques are 
adiabatic.}
\label{fig:evolution}
\end{figure*}

\Figure{fig:evolution} shows the evolution of our fiducial disk model.
The changes in density slope initially located at $\sim$0.4 and 21\,AU
correspond to opacity transitions (\citealp{Papaloizou}). The
constant temperature plateau initially at 4--5\,AU corresponds to the
opacity transition at $\sim$130\,K where ice grains sublimate (see
\Figure{fig:evolution}c). Initially, the ratio of radiative to dynamical timescales
$\varTheta>1$ through most of the disk ($\varTheta<1$ only at
$r>70$\,AU, beyond the model grid).  Thus, the torque in Equation~(\ref{eq:torque}) is mostly
adiabatic, and largely independent of the
interpolation procedure. 

Total disk mass and accretion rate decay nearly linearly with time. A
disk of 0.08 M$_{\odot}$ initially accreting at
$\ttimes{-7}\,M_{\odot}$~yr$^{-1}$ gets depleted in 4.8 Myr. At
that time, the total disk mass is $<1$\mearth, and we terminate the
simulation. The model gives a power-law decline of the density. 
An inner hole (\citealp{Clarke}) never forms, because of
the shallow radial slope of unity for the wind driven by external photoevaporation.

As the surface density drops, so does the viscous heating
rate (Equation~\ref{eq:temperature}), and the temperature falls accordingly,
down to $T=T_b$. During the evolution, the 
isothermal plateau 
shifts inwards from 4\,AU to 1\,AU by 4.5\,Myr. 
The optical depth only drops to unity at 10\,AU after 4\,Myr, 
and at 5\,AU after 4.4\, Myr, so the planet formation region is optically 
thick through most of the disk evolution. To check our method, we also ran models with 
Equation~(\ref{eq:temperature}),
and models where the temperature was derived using (1$+$1)-dimensional models, following
\citet{Papaloizou}. The 
differences were minor, consisting of a slightly larger isothermal plateau (extending to 6\,AU instead 
of 5\,AU), and an outer disk $\sim$10\,K hotter.
 
\begin{figure}
  \begin{center}
  \resizebox{.9\hfwidthsingle}{!}{\includegraphics{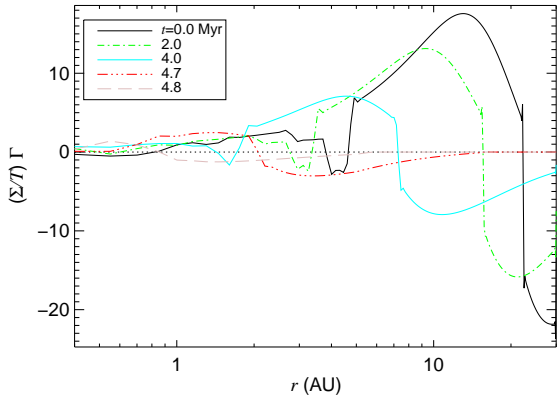}}
  \end{center}
\caption[]{Radial distribution of torque over time (Equation~\ref{eq:torque}).
Positive torques drive outward migration. The 
equilibrium radii of zero torque shift inwards as the disk 
evolves. The isothermal plateau corresponds to the well of 
negative torques initially at 4--5\,AU. Discontinuities 
correspond to opacity jumps. Torques are scaled by
$\varSigma/T$ to aid visualization.}
\label{fig:torques}
\end{figure}

To understand our results for planet migration in evolving disks, we 
first examine the behavior of planets in stationary models. In 
\Figure{fig:torques}, we show the torque as a function of radius. 
Migration halts at stable equilibrium radii where
$\Gamma=0$ and $d\Gamma/dr<0$, that is, where the torque is 
positive within (corresponding to outward migration) and
negative outside (inward migration). Two such equilibrium
radii occur,
at 4\,AU and 21\,AU, corresponding to the inner boundaries of the isothermal plateau and the outer
isothermal region, where $T=T_b$. (There is also a thin region
around an equilibrium radius at 0.6~AU that we do not consider here.)
A negative torque acts on planets migrating from 
$r_p>21$\,AU, bringing them to that radius, while a positive torque acts on 
planets with $4<r_p/{\rm AU}< 21$, also bringing them 
to that radius. Planets with $r_p<4$\,AU migrate to the inner
equilibrium radius. 

\begin{figure*}
  \begin{center}
    \resizebox{.9\hfwidth}{!}{\includegraphics{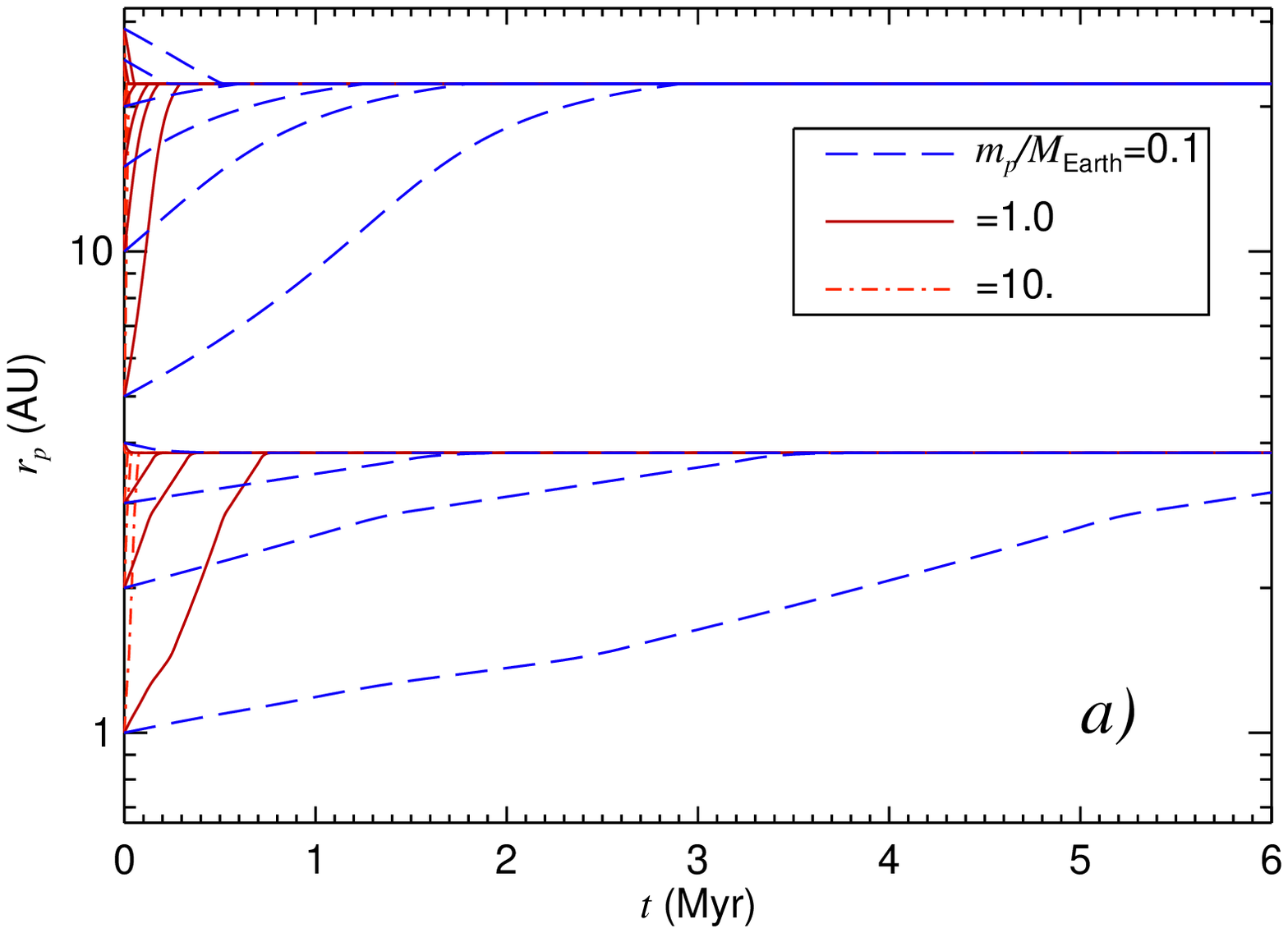}}
    \resizebox{.9\hfwidth}{!}{\includegraphics{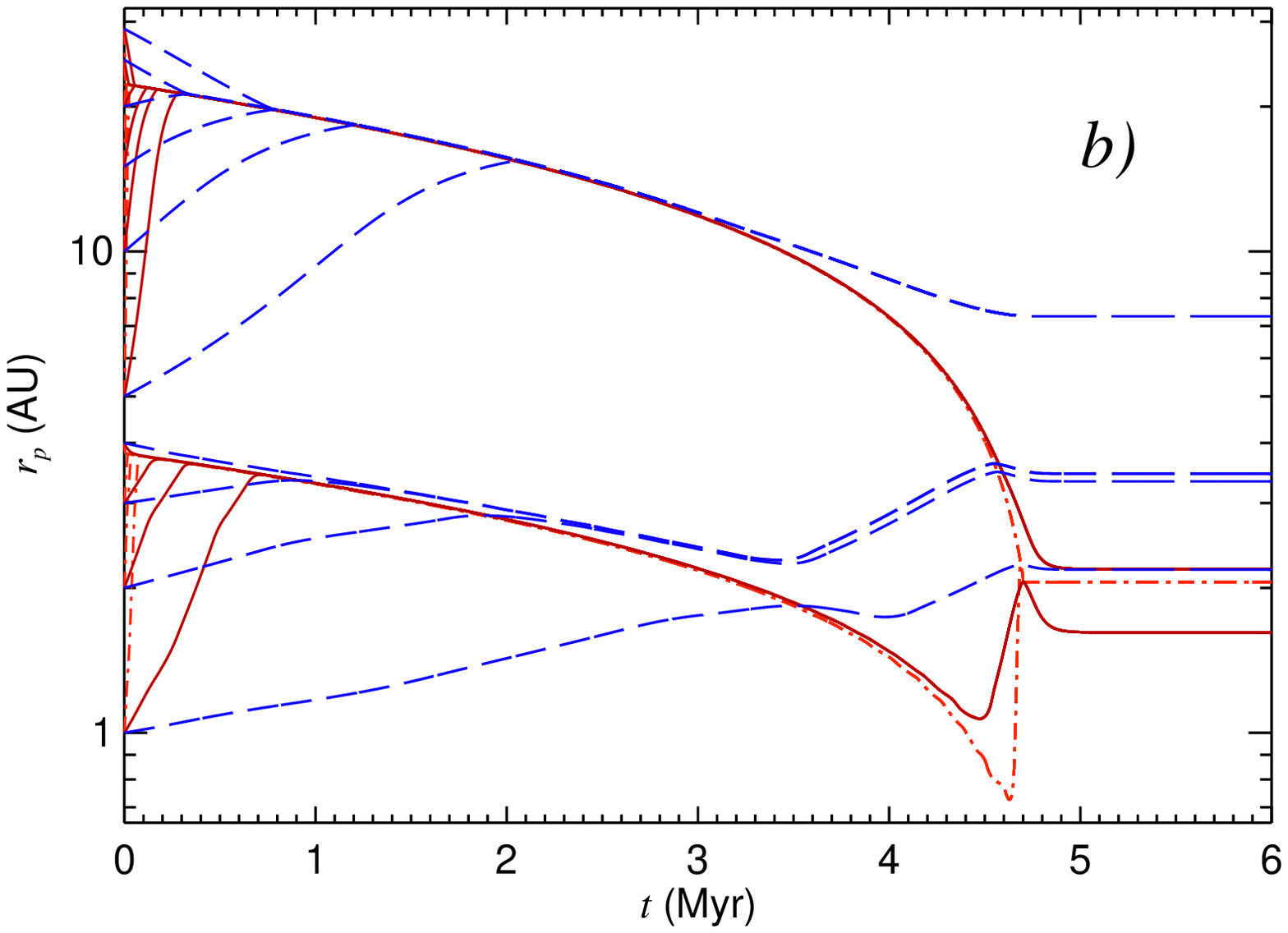}}
    \resizebox{.9\textwidth}{!}{\includegraphics{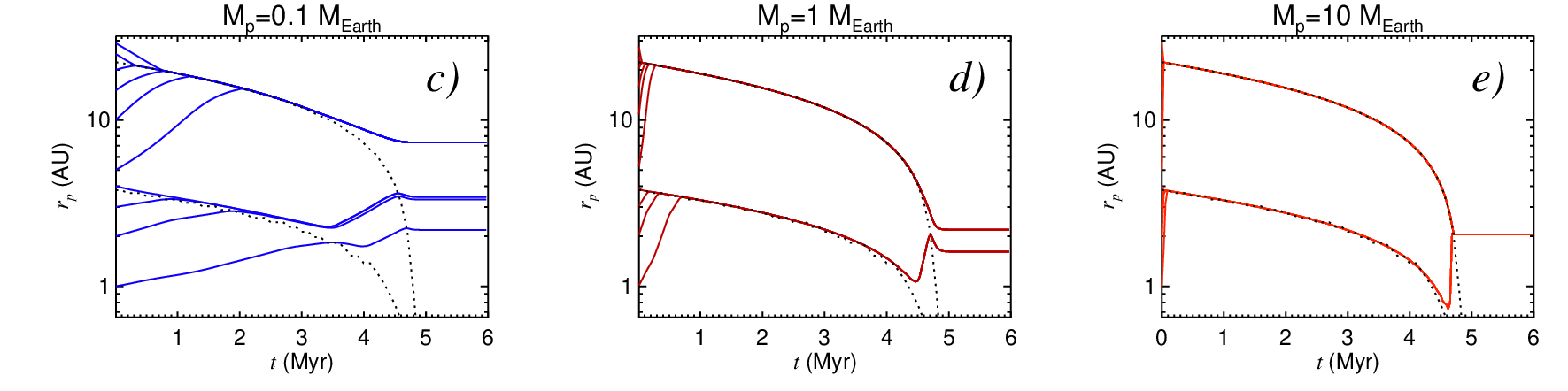}}
  \end{center}
  \caption[]{(a) Orbital migration in stationary disks. Planets
    migrate at mass-dependent rates toward equilibrium radii of zero torque (see
    \Figure{fig:torques}). 
    (b) Orbital migration in
    evolving disks.  Equilibrium radii migrate inwards on the
    accretion timescale, but the planets eventually migrate too slowly
    to remain in equilibrium. In the lower panels (c)--(e)
we show the same as in panel (b) but separated by planet mass.
The tracks of the equilibrium 
radii (dotted lines) are shown for comparison.}
  \label{fig:planet_gasevolution_linear}
\end{figure*}

We 
released planets of varying mass at several radii in the disk, and see precisely this behavior 
(\Figure{fig:planet_gasevolution_linear}a). 
The migration time to 
the equilibrium radius is inversely proportional to the 
planet mass, because the torque depends quadratically on mass
(Equation~\ref{eq:torque-norm}), so
$\dot{r}_p\propto{1/m_p}$ (Equation~\ref{eq:planet}). Planets of 10\,\mearthp  
 reach the 
equilibrium radii in $\leq0.1$\,Myr. In the isothermal 
case, planets migrate inwards at comparable speeds.

In \Figure{fig:planet_gasevolution_linear}b, we show trajectories of
planet migration in an evolving disk. Now 
the radii of zero torque shift inwards as the 
disk evolves. The planets migrate to these radii on the Type~I
timescale of $\ttimes{5}$\,yr, and then 
couple to disk evolution. Subsequently, they migrate inwards on the slow accretion  
timescale of $\ttimes{6}$\,yr, comparable to Type~II migration, as
predicted by \citet{Paardekooper10}.

\Figure{fig:planet_gasevolution_linear}b has some intriguing features that 
give pause. There is a clear effect of mass on the later evolution 
of the planets. In the stationary case, the only effect of mass is to 
determine the speed of migration to the equilibrium radii. In the
evolving case, the mass also 
helps determine the final location of the planet. To show this, we 
separately plot the tracks of planets of different masses in 
Figures~\ref{fig:planet_gasevolution_linear}c--e, as well as 
the trajectories of the equilibrium radii. The planets 
follow these equilibrium radii until late times, when they decouple. 

We can understand the process of decoupling by comparing the viscous
  accretion timescale to the migration timescale. As the disk thins, the surface density
  reaches a value so low that the gas cannot transfer
  sufficient angular momentum to the planet for its
  orbital radius to evolve as fast the equilibrium radius.
  Another way of understanding this is that a perturbation of the
  planet away from the equilibrium radius will only be corrected
  if the torque is sufficiently strong to return the planet
  before the radius moves a substantial distance. As the torque is a
  function of planet mass, the time and radius when decoupling occurs
  are also functions of the planet mass.

\begin{figure}
  \begin{center}
    \resizebox{.9\hfwidthsingle}{!}{\includegraphics{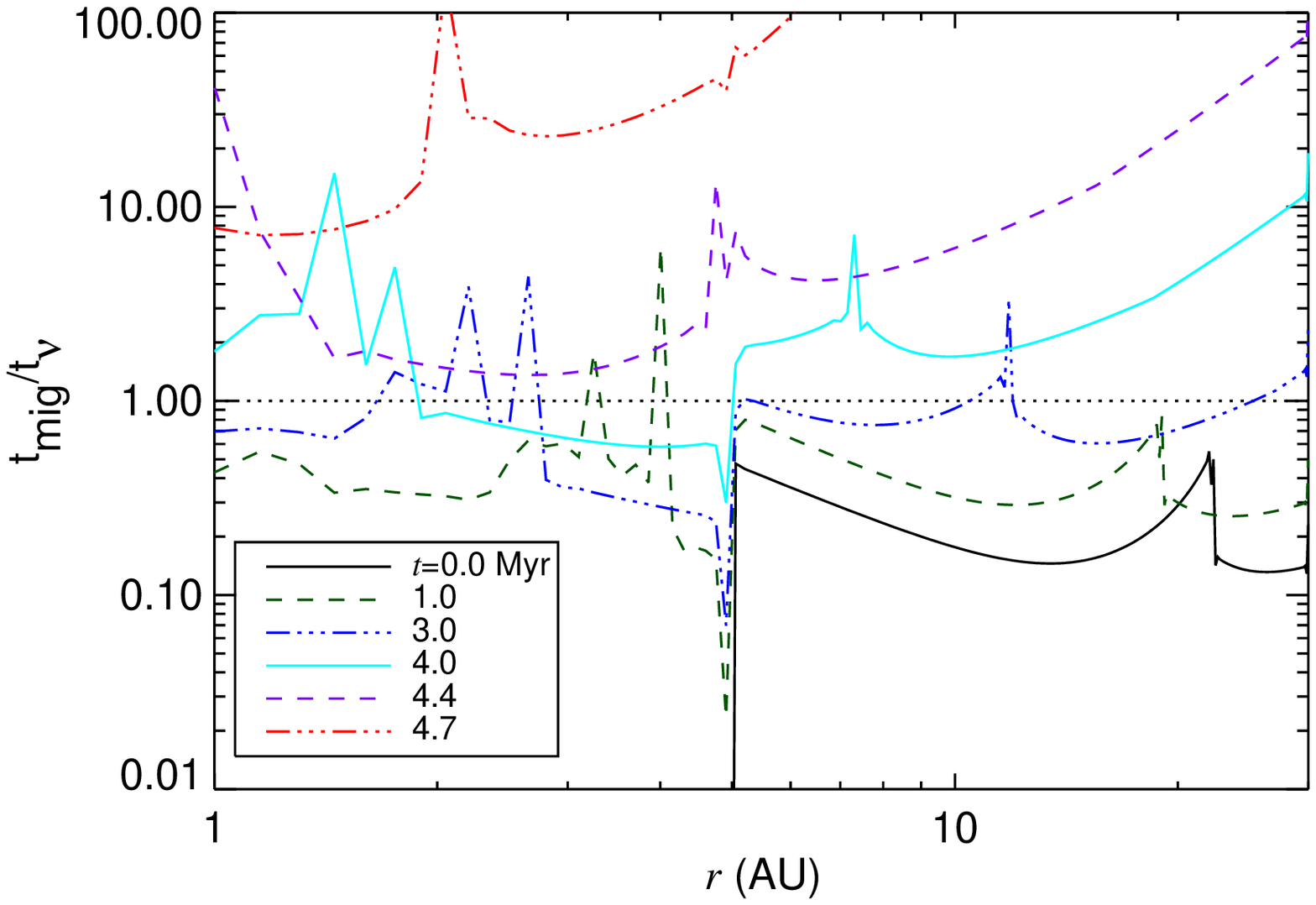}}
  \end{center}
\caption[]{Comparison of the timescales of planet migration ($t_{\rm mig}$) and 
disk accretion ($t_\nu$) for a planet of 1\mearth.} 
\label{fig:timescales}
\end{figure}

  We can estimate this time and radius by comparing the migration timescale
  ($t_{\rm mig}$=$r_p/|\dot{r}_p|$) to the disk accretion timescale
  ($t_{\nu}$=$\varSigma/|\dot{\varSigma}|$). As long as $t_{\rm mig}<
  t_{\nu}$, the planet can keep up with the evolution of the
  disk. When this is no longer true, 
decoupling occurs, releasing the planet.
We plot the quantity $t_{\rm mig}/t_{\nu}$ in \Figure{fig:timescales}, 
for the planet of 1\,\mearthp at different times. The structure in the plot 
comes from the radial derivatives of $\varSigma\nu$ that define the mass 
accretion flow, and from the torques $\Gamma$ that define the migration rate. 
The spikes in the figure correspond to the equilibrium locations, 
where $\Gamma=0$ (and therefore formally $t_{\rm mig}/t_{\nu}=\infty$). 
The decoupling in \Figure{fig:planet_gasevolution_linear}d starts to occur at 
$\approx4.0$\,Myr. In \Figure{fig:timescales}, we see that it roughly corresponds 
to the time when the vicinity of the equilibrium radius crosses the line of 
$t_{\rm mig}/t_{\nu}=1$. 

Evolution after decoupling proceeds as follows. The inner equilibrium
radius corresponds to the inner edge of the isothermal plateau. Once
the planet decouples, it is released inside the plateau, itself a
region of inward migration since the temperature gradient vanishes. As
disk evolution continues, the decoupled planet soon finds itself at
the outer edge of the isothermal plateau, and starts migrating
outwards. For planets of 1\mearth, decoupling occurs at $r_p$=1\,AU at
4.5\,Myr (\Figure{fig:planet_gasevolution_linear}b).  The planet then
rapidly descends the temperature gradient until it reaches the outer
equilibrium radius. However, this radius too moves inward faster than
the planet can migrate, so the planet enters the isothermal outer disk
(where $T\simeq{T_b}$), another region of slow inward migration. At
this stage, a planet of 1\,\mearthp still has time to migrate from
$r_p=2.0$~AU to $r_p=1.5$~AU before the density drops too low to drive
further migration (\Figure{fig:planet_gasevolution_linear}d).

The planet of 10\,\mearthp is strongly coupled and would follow the outer equilibrium 
radius until it hit the inner boundary of our model at 0.1\,AU. The halt at 2\,AU seen in 
\Figure{fig:planet_gasevolution_linear}e represents an artificial termination 
of the simulation. At that point, two criteria had been fulfilled. First, the scale 
height had become smaller than the Hill radius of the planet, so gap formation 
should have occurred. This is not sufficient to terminate the simulation, because 
the inward motion  of the equilibrium radius itself occurs at the timescale of 
Type II migration. However, the second criterion was that the mass parameter determining Type 
II migration, $\varSigma{r^2}$ (\citealp{Mordasini}), had become smaller than the 
planet's mass. At this stage, Type II migration becomes planet dominated and we consider 
that it comes to a halt. The smaller planets never carve gaps. 

We investigate migration in disks with different values of $\dot{M}_{\rm w}$ and $\dot{M_0}$, 
yet constrained by a lifetime of 1-10\,Myr,  without finding qualitative differences. 
A change in migration behavior is only seen for hotter disks. For 
$\alpha_{\rm SS}$=0.1, the disks show $\varTheta \approx 1$ 
throughout. Migration then shows a mix of isothermal and adiabatic behavior,
being mostly inwards yet with equilibrium points present. The torque only becomes 
isothermally dominated for $\alpha_{\rm SS}\simeq$1, which yields an unrealistically 
high accretion rate.

\section{Conclusions}
\label{sect:conclusions}

In this Letter, we examine the trajectories of planets undergoing 
Type I migration in evolving, radiative disks with initially nearly
adiabatic midplanes, where outward migration can occur.
Planets migrate toward equilibrium radii, where the torque 
acting on them vanishes. These radii correspond to opacity 
jumps and to the transition, in the outer disk, to an isothermal state. 
Because of viscous accretion and photoevaporation, these equilibrium radii 
themselves move inwards on disk accretion timescales. 

As long as the torques are strong enough to keep the planets coupled
to disk evolution, the planets migrate in lockstep with the gas at the
accretion timescale.  However, as the disk surface density drops, the
timescales of orbital migration and disk accretion eventually become
comparable. At this stage, if the planet is perturbed from an inner
equilibrium radius, the equilibrium radius moves inward faster than
the torques can return the planet, so it decouples.

If the continuing migration is outwards, the planet encounters another
equilibrium radius, and the same process of locking and decoupling
occurs. The outermost equilibrium radius lies where the disk reaches
the background temperature, and from there the planet can only migrate
inwards. However, by the time that the planet decouples even from that
equilibrium radius, the disk is already so severely depleted that the
ensuing inward migration is feeble, and soon comes to a halt, as the
remaining disk mass cannot exert a substantial torque. In no
case did a planet released beyond 1\,AU migrate all the way to the
star.

We stress that we only use a single model for the opacities,
which may change as planet formation progresses. The dust size
distribution must depend on the balance between coagulation and
fragmentation, which remains poorly understood. We also neglect
stellar irradiation, which will become important in the late stages of
the disk evolution. Irradiation will maintain high temperatures in the 
inner disk ($\approx$100\,K; \citealp{Chambers}), preventing gap 
formation for 10\mearthp planets. On the other hand, irradiation should 
lead to a hole-forming photoevaporating wind (\citealp{Clarke,Alexander}), 
which quickly depletes the disk, thus possibly 
bringing migration to an even earlier halt. Future work should 
self-consistently address these issues.

As a consequence of the independence of equilibrium radii on
planet mass, all planets migrate to these equilibrium
locations. Ensembles of planets reaching them may become
violently unstable due to $N$-body interactions. Nevertheless, even if
scattered away, migration will invariably drive the planets back toward
these radii. The final outcome may well be collisions driving
further planet growth, aiding rapid giant planet formation
or forming planets in 1:1 resonance. 

If this is the case, however, it raises the question of why the solar
system has a set of neatly spaced planets as opposed to only two, as
the two equilibrium radii of the model might naively suggest. One
possible solution is that $\Gamma$ (and thus any equilibrium radius)
shows a dependence on the planet-to-star mass ratio $q$ at the verge
of gap opening when $q\approx{h^3}$ (\citealp{Masset}). Another is
that we only consider the fully unsaturated torque, whereas saturation
depends on the width of the horseshoe region and therefore on the
planet's mass. However, the level of saturation in radiative disks is
not fully understood at present, and we cannot easily add it to our
study. Future models should include effects of saturation to study possible mass segregation.
Finally, in view of the long migration timescales for
$M\lesssim{0.1}$\mearth, such planets may just not have the time to
migrate back to the equilibrium location before the disk vanishes if
scattered far enough.  This scattered population of small planets
could provide the initial conditions for the terrestrial planets
of our own solar system.

Our results provide qualitative and quantitative justification for 
the reduction of Type I migration rates assumed in planetary population 
synthesis models (\citealp[e.g.,][]{Alibert,Ida,Mordasini}). 
Instead of migrating on the fast, mass-dependent, 
timescale $t_{\rm mig}$, we find that planets spend their first Myr near
equilibrium radii that change only on the slow accretion timescale
$t_{\nu}$. We show in \Figure{fig:timescales} that $t_{\rm mig}/t_{\nu}\sim0.1$ 
for a 1\mearthp planet during most of the evolution of the disk. Examining 
the same figure for different masses shows a linear dependence on mass, $t_{\rm mig} /
t_{\nu} \sim 0.1 (M / \mbox{\mearth})$, consistent with the population 
synthesis assumptions.

\acknowledgments W.L. is partly supported by a Kalbfleisch Research
Fellowship from the AMNH. M.-M.M.L. is partly supported by NSF CDI grant AST-0835734,
and NASA OSS grant NNX07AI74G. S.-J.P. is an STFC postdoctoral fellow. We acknowledge useful 
discussions with C. Mordasini and T. Birnstiel.

\end{document}